\documentclass{aa}
\usepackage{natbib,graphicx}
\bibpunct{(}{)}{;}{a}{}{,}
\def\be {\begin{equation}}
\def\ee {\end{equation}}
\def\simless{\mathbin{\lower 3pt\hbox
    {$\rlap{\raise 5pt\hbox{$\char'074$}}\mathchar"7218$}}} 
\def\simgreat{\mathbin{\lower 3pt\hbox
    {$\rlap{\raise 5pt\hbox{$\char'076$}}\mathchar"7218$}}} 
\def\msun {$M_\odot$}
\def\mic {$\mu\hbox{m}$}

\def\percc {$\hbox{{\rm cm}}^{-3}$}    
\def\cmsq  {$\hbox{{\rm cm}}^{-2}$}    
\begin{document}


\title{The structure and stability of molecular cloud cores \\
in external radiation fields}

\author{Daniele Galli\inst{1} \and Malcolm Walmsley\inst{1} \and Jos\'e Gon\c calves\inst{1,2}}

\institute{INAF-Osservatorio Astrofisico di Arcetri, Largo E. Fermi 5, I-50125 Firenze, Italy \and
Centro de Astronomia e Astrof\'{\i}sica da Universidade de Lisboa, Tapada da Ajuda, 1349-018 Lisboa, Portugal}

\offprints{M.~Walmsley, \\
\email{walmsley@arcetri.astro.it}}

\date{Received / Accepted}

\authorrunning{Galli, Walmsley \& Gon\c calves}
\titlerunning{Cloud cores in external radiation fields}

\abstract{
We have considered the thermal equilibrium in pre-protostellar cores
in the approximation where the dust temperature is independent of
interactions with the gas and where the gas is heated both by collisions
with dust grains and ionization by cosmic rays. We have then used these
results to study the stability of cores in hydrostatic equilibrium
in the limit where thermal pressure dominates over magnetic field
and turbulence.  We compare the density distribution derived in this
manner with results obtained in the isothermal case.  We find that for
cores with characteristics similar to those observed in nearby molecular
clouds, the gas and dust temperatures are coupled in the core interior
with densities above $\sim 3\times 10^4$~cm$^{-3}$.  As a consequence,
one expects that the  gas temperature like the dust temperature decreases
towards the center of these objects. However, the regime where gas and
dust temperatures are coupled coincides approximately with that in which
CO and many other molecular species deplete onto dust grain surfaces.
At larger radii and lower densities, the gas and dust temperatures
decouple and the gas temperature tends to the value expected for cosmic
ray heating alone.  The density structure which one computes taking into
account such deviations from isothermality are not greatly different from
that expected for an isothermal Bonnor-Ebert sphere.  It is impossible
in the framework of these models to have a stable equilibrium core with
mass above $\sim 5$~\msun\ and column density compatible with observed
values ($N_{\rm H} > 2\times 10^{22}$~\cmsq\ or $A_{\rm V} > 10$ mag.). We
conclude from this that observed high mass cores are either supported by
magnetic field or turbulence or are already in a state of collapse. Lower
mass cores on the other hand have stable states where thermal pressure
alone provides support against gravitation and we conclude that the
much studied object B68 may be in a state of stable equilibrium if the
internal gas temperature is computed in self-consistent fashion. Finally
we note that in molecular clouds such as Ophiuchus and Orion with high
radiation fields and pressures, gas and dust temperatures are expected
to be well coupled and hence in the absence of an internal heat source,
one expects temperatures to decrease towards core centers and to be
relatively high as compared to low pressure clouds like Taurus.
}

\maketitle

\section{Introduction}

It has long been presumed that during the formation of a star, there is
an intermediate phase in which the ``protostar'' is at least approximately
in a state of hydrostatic equilibrium or magneto--hydrostatic equilibrium
\citep[see e.g.][]{sal87}.  While this idea in origin was merely based on
plausibility arguments, it has received support from the discovery that in
least some nearby molecular clouds, one finds embedded ``cores'' of higher
density than the surroundings where the observed linewidths are thermally
dominated.  That is to say, while there may be local subsonic turbulence,
there is no evidence for collapse onto a point mass and it appears that
thermal pressure is capable of balancing gravity.  On the other hand,
comparisons of the gravitational, thermal, magnetic and turbulent energies
of such cores show that all these quantities are equal to within the
(considerable) uncertainties \citep[][]{mg88a,mg88b,mal91,c99}. These
data suggest though they do not prove the existence of equilibrium
structures which are an intermediate state in the evolution of a {\it
prestellar core}.

This picture has received more observational support with the advent of
high quality maps of the millimeter continuum emission of the dust grains
within such cores as well as the possibility to study their extinction
in the near and mid infrared \citep{jwm2000,jfmm2001,bap00,llbc94}.
These have allowed a more unbiased view to be obtained of the
density distribution in prestellar cores. In particular, they
have shown that the early molecular line data was highly biased
because above a critical density of $\sim 5\times 10^4$ cm$^{-3}$
\citep[see][]{tmc02,cwz02a,cwz02b,kal99} most molecular tracers
including CO condense out onto dust grain surfaces.  This depletion has
the consequence that in molecular lines one sees mainly a lower density
outer shell, whereas the dust emission (or absorption) offers a more
unbiased view of the density distribution.

The results from the millimeter continuum and infrared absorption studies
have been compared with a variety of theoretical models of hydrostatic
cores \citep{bap00}. One result of such studies has been evidence for a
``flattening'' in the density distribution for radii below a critical
value of $r_{\rm cr}\simeq 2000$-8000 AU implying a steeper fall off in
radius above $r_{\rm cr}$. Thus for example, \citet{bap00} model L1544
with a roughly uniform density inside 1900 AU but a rapid fall off
outside this radius.  Structures  whose support against gravitational
collapse is mainly due to the magnetic field are plausible both because
the observed dust continuum maps show large departures from spherical
symmetry and because of time scale arguments.  Collapse on a free fall
time scale would produce a larger star formation rate than that observed.

There are however some cores where thermal pressure may dominate
magnetic pressure and spherical symmetry may be a good assumption.
Particularly worthy of note is B68 where \citet{al01} have recently
demonstrated that the density distribution derived from their NIR
measurements is consistent with a purely thermally supported hydrostatic
model. In fact, they find results consistent with the equilibrium
structures discussed by \citet{e55} and \citet{b56}, which we will
refer to in the following as Bonnor-Ebert spheres. To what extent B68
is exceptional is presently unclear but we note that a large number
of cores in the Ophiuchus and Orion clouds \citep{jwm2000,jfmm2001}
appear to have characteristics compatible with Bonnor-Ebert spheres at
temperatures of 15--30~K.

This accord between theoretical expectation and observation suggests
that a study of the theoretical assumptions may be worthwhile.  One of
these assumptions has been that of isothermality which is often based
on the concept of ``low optical depth for the cooling transitions''.
In reality, the gas in prestellar cores is thought to be cooled
mainly by optically thick CO transitions \citep{g01} although, as mentioned
above, the CO seems to disappear at high densities.  Moreover, the
typical density for which dust grains and gas become thermally coupled
is roughly of the same order as that observed in prestellar cores
\citep[see, e.g.][]{kw84}.  Thus, we decided that a new look at the gas
temperature distribution to be expected in such cores seemed
warranted.  

In this paper we replace the assumption of an isothermal gas with the more
realistic condition of thermal balance in the gas, and we evaluate the
consequences of an external heating source (the interstellar radiation
field) on the structure of the cloud (density and temperature profiles)
and its stability properties.  Recent studies of the dust temperature
distribution in such objects \citep{eal01,zwg01} have shown that the dust
temperature typically falls by a factor of 2 from edge to center. It
seems reasonable to ask how the gas temperature will react in such
circumstances. One might also ask  whether the density distribution in
hydrostatic equilibrium will depart appreciably from that expected under
the isothermal assumption. Will the expected density contrast differ for
example from that expected for a Bonnor-Ebert sphere when one calculates
the gas temperature in self--consistent fashion?  This article represents
an attempt to answer such questions.

The outline of this paper is as follows. In Sect.~2 we give a brief
introduction to the theory of structures in hydrostatic equilibrium
including the results for an isothermal equilibrium Bonnor-Ebert
sphere. In Sect.~3 we discuss the input to our calculations and the
simplifications which we have made. In Sect.~4, we present our results
for the gas temperature distribution in two model cores whose density
distribution has been assumed similar to that observed in L1544 and
B68. In Sect.~5 we present our results for non-isothermal hydrostatic
equilibria for a variety of conditions.  Here we show among other things
that the equilibria obtained depend relatively sensitively on the external
radiation field. In Sect.~6 we discuss the observational consequences
of our results, and compare the properties of our model clouds
to isentropic polytropes. In Sect.~7 we summarize our conclusions.

\section{Theoretical background}

An isothermal gas sphere embedded in an external medium of given
pressure $p_{\rm ext}$ has a critical Bonnor-Ebert mass above which no
state of equilibrium can exist \citep{e55, b56, mc57},
\begin{eqnarray}
\lefteqn{M_{\rm BE}\simeq 1.182 \frac{a^4}{\sqrt{G^3p_{\rm ext}}}} \nonumber \\
& & \simeq
2.6~\left(\frac{T_{\rm g}}{10~\mbox{K}}\right)^2
\left(\frac{p_{\rm ext}}{2\times 10^4~\mbox{K}~\mbox{cm}^{-3}}\right)^{-1/2}M_\odot,
\label{mbe}
\end{eqnarray}
where $a=(kT_{\rm g}/\mu m_{\rm H})^{1/2}$ is the sound speed in the
gas, $T_{\rm g}$ the gas temperature and $\mu$ the mean molecular
weight (see Fig.~\ref{fig_m1}).

Below this critical value, the equation of hydrostatic equilibrium admits
single or multiple solutions, characterized by different degrees of
density concentration. It has been shown by \citet{b56} and \citet{e57}
that equilibrium configurations with center-to-boundary density contrast
$\rho_{\rm c}/\rho_{\rm b} < 13.98$ are stable, whereas equilibrium
solutions with $\rho_{\rm c}/\rho_{\rm b} > 13.98$ are unstable to
radial collapse.  Stable isothermal equilibria can also be characterized
by the condition
\be
\xi_{\rm max}\equiv \frac{R}{a}\sqrt{4\pi G\rho_{\rm c}}<6.451,
\ee
where $R$ is the cloud's radius.

Recently, \citet{lb01} have re-examined the classical Bonnor-Ebert
problem relaxing the assumption of spherical shape of the cloud. The
result is that for non-spherical isothermal clouds, the critical
Bonnor-Ebert mass is {\it larger} than the value given by
Eq.~(\ref{mbe}); a sphercal shape is more prone to gravitational
instability than clouds of other shapes.  The density contrast
$\rho_{\rm c}/\rho_{\rm b}$ for marginally stable non-spherical clouds
is different from the Bonnor-Ebert value 13.98 for spherical clouds,
but, remarkably, the condition for stability, when expressed in terms
of the ratio of the average density $\bar\rho$ to the boundary density
$\rho_{\rm b}$, i.e. $\bar\rho/\rho_{\rm b} < 2.465$, is independent of
cloud shape.

To express the condition for stability ($M<M_{\rm BE}$ and $\rho_{\rm
c}/\rho_{\rm b} < 13.98$) in terms of observable quantities one can
eliminate the external pressure writing $p_{\rm ext}=a^2\rho_{\rm b}>
13.98^{-1}a^2 \rho_{\rm c}$, obtaining, for a given cloud's mass $M$,
\be
\rho_{\rm c} < 19.53 \frac{a^6}{G^3 M^2},
\label{rhocrit}
\ee
that, for a cloud of molecular hydrogen with mean molecular weight
$\mu=2.33$ becomes
\be
n_{\rm c}<1.6\times 10^5 \left(\frac{T_{\rm g}}{10~{\rm K}}\right)^3
\left(\frac{M}{M_\odot}\right)^{-2}~\mbox{cm$^{-3}$}.
\label{ncrit}
\ee
Thus, {\it stable} isothermal equilibria are only possible provided the
central density is relatively low (in the absence of magnetic fields):
this causes a conflict with observations if one attempts to model
observed molecular cores with such equilibrium configurations. For
example, if dense cores with $M > 4$~\msun\ and $T_{\rm g}=10$~K are
modeled as stable Bonnor-Ebert spheres, their density according to
Eq.~(\ref{ncrit}) will fall below the critical value of $\sim 10^4$
\percc \ reqired to excite the molecular lines observed in these regions.
Moreover, the central densities inferred in many pre--stellar cores based
on the observed mm--submm dust emission \citep[e.g.][]{wma99} is often
above $10^5$ \percc \ and hence for gas temperatures of $\sim 10$~K,
only cores of mass below 1~\msun\ can be in stable equilibrium.

Many of the observed objects however have higher masses (see the
discussion of \citet{bap00} and also column densities of up to $10^{23}$
\cmsq \ in molecular hydrogen or 100 visual magnitudes of extinction. One
can express the stability condition for Bonnor-Ebert spheres as a
condition on the cloud's total (from edge to edge) extinction through
the center $A_{\rm V,tot}=N({\rm H}_2)/10^{21}~{\rm cm}^{-2}$, obtaining
\be
A_{\rm V,tot}\simless 29 \left(\frac{T_{\rm g}}{10~{\rm K}}\right)^2
\left(\frac{M}{M_\odot}\right)^{-1}~\mbox{mag}.
\ee
This is clearly much less than often observed and provides a strong
argument for the support of many observed structures being due to
forces other than pure thermal pressure \citep[though there are also
cases of thermal support, e.g.][]{al01,jwm2000,jfmm2001}. In the
following, we consider ``observed cores'' as having a minimum $A_{\rm
V}$ of 10 mag., and thus exclude processes such as UV photoelectric
effect from consideration.  We do not moreover in this work compute
models including magnetic fields and turbulence, but briefly mention in
the next Sections how these agent of support modify the picture outlined
above.

In summary, non-magnetic stable isothermal cores represent only
moderate enhancements (a factor $\sim 2.5$ in the average
density) of the ambient gas density, quite independent on cloud shape
and the turbulent field. This limited density range represents a
serious limitation when modeling molecular cloud cores as
non-magnetized hydrostatic equilibria. For example \citet{bdj84} were
able to reproduce column density, brightness temperatures and
abundances of several chemical species measured in dark clouds like
L134, L183 and TMC-1 only with models characterized by
density contrast $\rho_{\rm c}/\rho_{\rm b}\simeq 2000$. \citet{h88}
however verified that these models were well into the unstable regime.
The same criticism was made by \citet{cpdf87} to the hydrostatic models
computed by \citet{fp85}. \citet{h88} concluded that {\it stable}
spherical, hydrostatic models able to satisfy all observational
constraints on extinction, external pressure, mass, and chemical
composition of molecular cloud cores, could no be constructed. In this
respect, it is interesting to recall that when molecular cloud cores
are fitted by Bonnor-Ebert spheres, the best fit is obtained for models
that are in the unstable regime: $\xi_{\rm max}=6.9$ for
B68~\citep{al01}, $\xi_{\rm max}=12.5$ for B335~\citep[][although B335
is not a starless core]{har01}.

\subsection{Magnetic fields}

The additional support provided by a large-scale
magnetic fields can be evaluated by a virial-theorem analysis or with
the help of detailed calculations \citep[see e.g.][ for a
review]{mkal93}. The maximum value of a cloud mass for the existence
of stable equilibria is approximately equal to
\be
M_{\rm cr}\simeq \left[1-\left(\frac{0.13}{\lambda}\right)^2
\right]^{-3/2}M_{\rm BE},
\ee
where $\lambda$ is the magnetic mass-to-flux ratio of the cloud in
units of $G^{-1/2}$ \citep{ms76,tin88}.  For magnetically subcritical
clouds ($\lambda\simless 0.13$), no amount of external compression can
induce the collapse of the cloud (as long as the magnetic field remains
frozen in the matter).  Magnetically supercritical clouds
($\lambda\simgreat 0.13$) can exist in stable equilibrium for masses
larger than the Bonnor-Ebert critical mass, and can reach values of
$\rho_{\rm c}/\rho_{\rm b}$ much higher than $13.98$ \citep[up to
$\sim 100$, see Fig.~4 and 6 of][]{tin88}.

\subsection{Non-uniform turbulence and temperature gradients}

If the turbulent motions of the gas are characterized by a uniform
(three-dimensional) mean square velocity $\langle {\bf v}_{\rm turb}^2
\rangle$, the effects of a turbulent pressure, in addition to the
thermal pressure, can be trivially incorporated in the classic
Bonnor-Ebert picture summarized above.  The condition for the existence
of stable equilibria is the same expressed by Eq.~(\ref{mbe}) with the
sound speed $a^2$ replaced by an ``effective'' sound speed $a_{\rm
eff}^2=a^2+\langle {\bf v}_{\rm turb}^2 \rangle/3$, allowing the
existence of stable equilibria for values of the central density (or
peak extinction) larger than for clouds supported by thermal pressure
alone.  Notice, however, that (as in the purely thermal case)
marginally stable clouds are characterized by a density contrast
$\rho_{\rm c}/\rho_{\rm b} = 13.98$ or $\bar\rho/\rho_{\rm
b}=2.465$, where $\bar\rho$ is the average density,
irrespectively on the cloud's {\it effective} temperature. 

One should remember also that the observed line widths (which in some
cases are close to being thermal) place limits on possible values of
${\bf v}_{\rm turb}$.  In dense cores, as well as in giant molecular
clouds, turbulent linewidths are often observed to increase at larger
scales \citep[see discussion in][]{bg98,gbwh98}, suggesting that these
objects can be modeled as negative-index polytropes \citep{djdb80,
dc83, m88}.  This is easily understood comparing the polytropic
equation of state $P\propto \rho^{1+1/n}$ with the expression for the
thermal pressure of a gas, $P\propto \rho T$, implying $T\propto
\rho^{1/n}$. Since the density is always a decreasing function of
radius, negative index polytropes represent equilibrium configurations
in which the effective temperature increases outward. However, from the
point of view of the stability properties, our models show a behavior
similar to polytropes with $n\gg 1$.  We will return to this point in
Sect.~6.2.

The logatropic equation of state analyzed by \citet{mlp96} to incorporate
the contribution of on non-thermal motions to the support of molecular
clouds (and cloud cores) makes possible the existence of stable equilibria
for large density constrasts ($\rho_{\rm c}/\rho_{\rm b}\simeq 100$) but
the resulting configurations are much more extended than Bonnor-Ebert
spheres and are characterized by a maximum ratio $\bar\rho/\rho_{\rm
b}=3/2$, even smaller than for critical Bonnor-Ebert spheres. Indeed,
the observed column density distributions \citep{bap00} of cores do not
fit well to logatropes.

\section{Model assumptions}

Following the approach outlined by ~\citet{g01}, we compute the gas
thermal equilibrium assuming that the dust temperature is unaffected by
collisions with gas particles \citep[see][for estimates of the validity
of this]{eal01}.  Moreover, we have used the analytical fit of
\citet{zwg01} to determine dust temperature as a function of radius in
our spherically symmetric model cores.  The justification for this is
simplicity. We are interested here in understanding qualitatively to
what extent departures from isothermality may affect the properties of
pre-stellar cores.  High precision in determining the gas cooling rates
and the dust temperature distribution is unlikely to fundamentally
affect our results.

\subsection{Dust and gas temperature}

Thus we determine the gas temperature $T_{\rm g}$ by solving
the equation of thermal balance of the gas in the presence of dust
grains heated by the external radiation field,
\be
\Gamma_{\rm cr}-\Lambda_{\rm g}-\Lambda_{\rm gd}=0,
\label{balance}
\ee
where $\Gamma_{\rm cr}$ is the cosmic-ray heating rate, $\Lambda_{\rm
g}$ the gas cooling rate by molecular and atomic transitions, and
$\Lambda_{\rm gd}$ the gas-dust energy transfer rate.
We follow \citet{g01} and adopt a cosmic-ray heating rate
\be
\Gamma_{\rm cr}=10^{-27}
\left(\frac{\zeta}{3\times 10^{-17}~\mbox{s$^{-1}$}}\right)
\left[\frac{n({\rm H}_2)}{{\rm cm}^{-3}}\right]~{\rm erg~cm}^{-3}~{\rm s}^{-1}.
\label{crheat}
\ee
The gas cooling rate $\Lambda_{\rm g}$ for typical conditions of molecular
cloud cores has been computed by \citet{g01} in the LVG approximation
(assuming a velocity gradient of 1~km~s$^{-1}$~pc$^{-1}$) for various
degrees of depletion of molecules from the gas phase onto grain surfaces.
We adopt the parametrization of \citet{g01}
\be
\Lambda_{\rm g}=\alpha
\left(\frac{T_{\rm g}}{10~{\rm K}}\right)^\beta~{\rm erg~cm}^{-3}~{\rm s}^{-1},
\label{gascool}
\ee
where $\alpha$ and $\beta$ are parameters that depend on the H$_2$
density and the molecular depletion factors. When not explicitly noted,
we assume that molecular abundances in the gas have their standard
(undepleted) values.  As for the the gas-dust energy transfer rate we
adopt the expression given by \citet{g01} based on calculations of
\citet{bh83},
\begin{eqnarray}
\lefteqn{\Lambda_{\rm gd}=2\times 10^{-33}
\left(\frac{n}{{\rm cm}^{-3}}\right)^2
\left(\frac{T_{\rm g}-T_{\rm d}}{{\rm K}}\right)} \nonumber \\
& &
\times \left(\frac{T_{\rm g}}{10~{\rm K}}\right)^{1/2}~{\rm erg~cm}^{-3}~{\rm s}^{-1}.
\label{gasdust}
\end{eqnarray}

To determine the dust temperature $T_{\rm d}$ we ignore the gas-dust
coupling and we solve the equation of thermal balance of the dust
\be
\Gamma_{\rm ext}-\Lambda_{\rm d}=0,
\ee
where $\Gamma_{\rm ext}$ is the dust heating rate by the external
radiation field and $\Lambda_{\rm d}$ the dust cooling rate by
radiation. In practice, we use the analytical solution of \citet{zwg01}
to compute $T_{\rm d}$ as function of extinction from the cloud's
boundary.

As in \citet{zwg01}, we have adopted as a standard the interstellar
radiation field (ISRF) in the solar neighborhood given by \citet{mmp83}
and \citet{b94}. It consists of four components: the V-NIR component,
peaking at $\lambda\simeq 1$~\mic, the MIR and FIR components, peaking
at $\lambda\simeq 100$ and 140~\mic\ respectively, and the cosmic
background radiation, peaking at $\lambda\simeq 1$~mm. We allow for
variations of the ISRF, assuming, for simplicity, that the spectral
shapes of these components are the same as the local ISRF, and we scale
the intensity of the V-NIR, MIR, and FIR components by a factor $G_0$
(with $G_0=1$ for the local ISRF).

The existence of a cloud radius leads to the necessity of an external
pressure to keep the system in equilibrium. We embed our model cloud
cores in a spherical envelope that represents the ambient molecular
cloud. We do not need to specify the physical properties of this
envelope; we just assume that it provides the needed external pressure
and a shielding from the ISRF directly incident on the cloud cores. We
set the latter at the value $A_{\rm V}^{\rm env}=1$ \citep{mk99},
sufficient to absorb the UV and most of the V-NIR radiation. We define
$A_{\rm V,c}$ to be the centre-to-edge extinction of our model cloud
cores. The range of masses and densities of
interest here corresponds to the range $5\,\simless A_{\rm V,c}\,\simless 100$,
where the central dust temperature is determined mostly by optically thick
absorption of V-NIR radiation and optically thin absorbtion of MIR and
FIR radiation.

The fundamental parameters of our models are the external pressure
$p_{\rm ext}$ and the intensity of the external radiation field $G_0$.
In this paper, we consider independent variations of these two
quantities, and we analyse separately the dependence of the critical
mass for increasing $p_{\rm ext}$ and fixed $G_0$ (Sect.~5,
Fig.~\ref{fig_m1}), and the effect of varying $G_0$ for a given value
of $p_{\rm ext}$ (Sect.~6, Fig.~\ref{fig_static}).

\subsection{Effects of molecular depletion}

We include in our models the effect of molecular depletion
onto dust grains, parametrized for simplicity by the single parameter
$f_{\rm d}$ (depletion factor), defined as the ratio of the ``standard''
abundance of CO isotopes and CS to the actual abundances.
The corresponding abundances of other molecular species are given in
Table~3 of \citet{g01}. Gas-phase cooling rates for several values of
$f_{\rm d}$ in the range 1--100 were computed and tabulated by \citet{g01} for
H$_2$ density from $10^3$~cm$^{-3}$ to $10^6$~cm$^{-3}$.  We assume the
dependence of $f_{\rm d}$ on density given by \citet{tmc02},
\be
f_{\rm d}=\exp\left(\frac{n}{n_{\rm dep}}\right),
\ee
where the critical density for CO depletion is taken to be 
$n_{\rm dep}=5.5\times 10^4$~\percc.  


Although we are mainly interested in applications to molecular cloud
cores of moderate to high extinction, where molecular depletion is
effective only in a region of limited spatial extent, one should keep
in mind that large fraction of the cloud's mass is contained in the
outer parts (roughly $M(r)\propto r$), and therefore the overall
stability of the cloud may be modified by the thermal properties of
these external layers.

In our models we find in general that the effects of depletion on the
temperature of the gas are significant only for densities $n \simless
10^5$~cm$^{-3}$. At higher density, gas-dust coupling overwhelms the
reduction of the cooling rate of the gas, and the gas temperature
becomes insensitive to depletion.  This is in qualitative agreement
with the results of \citet{g01}.  We conclude with \citet{g01} that
molecular depletion in the central, dense regions of molecular cloud
cores ($n\simgreat 10^{4.5}$--$10^5$~cm$^{-3}$) is not associated with
an increase in the gas temperature.

In the extreme case where CO (and other
molecules containing C,N,O,S, etc.) depletes out onto grains causing
Eq.~(\ref{balance}) to reduce to $\Gamma_{\rm cr} = \Lambda_{\rm gd}$. 
Then using Eq.~(\ref{crheat}) and (\ref{gasdust}), one finds
\begin{eqnarray}
\lefteqn{\frac{T_{\rm g}-T_{\rm d}}{T_{\rm g}} \simeq} \nonumber \\
& & 5\left(\frac{\zeta}{3\times 10^{-17}~\mbox{s}^{-1}}\right)
\left(\frac{n}{10^4~\mbox{cm}^{-3}}\right)^{-1}
\left(\frac{T_{\rm g}}{10~\mbox{K}}\right)^{-1.5},
\end{eqnarray}
This shows that in the extreme case where all heavy species deplete
out, coupling between gas and dust temperatures will be assured for
densities larger than $5\times 10^4$~\percc\ if the cosmic ray
ionization rate is standard or lower. Since it is essentially at
densities above this value that CO is observed to deplete out, we
conclude that a rough equality between gas and dust temperatures in the
depleted region is a reasonable assumption. In this context, it is
interesting to note that \citet[][]{bal02} suggest that even such
relatively volatile species as molecular nitrogen have depleted out in
the central core of B68.

\section{Temperature distributions for L1544 and B68}

Before considering what one expects for theoretical equilibrium cores,
it is useful to consider what one predicts for the temperature
variation within an object whose density structure is similar to those
observed. Specifically, we will consider the case of L1544
\citep{cwz02a,cwz02b} and B68 \citet{al01}, two well studied 
prestellar cores.

To model L1544 we assume $G_0=1$ and the density structure of \citet{tmc02}, 
which, in turn, is based on the 1.3mm dust emission and absorption maps of
\citet{bap00}.  Thus, we assume the H$_2$ density distribution $n(r)$ to be
given by
\be
n(r)=\frac{n_{\rm c}}{1+(r/r_0)^{2.5}},
\label{l1544den}
\ee
where the central ${\rm H}_2$ density is $n_{\rm c}=1.4\times
10^6$~\percc\ and $r_0=0.014$~pc. It is worth noting here that we have
on purpose taken here a somewhat extreme case with large density
contrast relative to the surroundings and large central column density
\citep[$6.0 \times 10^{22}$ \cmsq\ corresponding to 60 mag. of
extinction, from][]{bap00}. One should also be aware that the aspect
ratio of the core based on the dust emission contours is $\sim 0.6$ and
hence spherical symmetry is a crude assumption. Nevertheless, the case
is of interest in that it is a model which approximates the actually
observed density distribution and in which, one expects that the
central temperature will be relatively low (due to the large
extinction).

For B68 we assume the density structure of a Bonnor-Ebert sphere with
the physical parameters given by \citet{al01}.  The central density
$n_{\rm c}=2.4\times 10^5$~cm$^{-3}$ is a factor $\sim 6$ lower than in
the case of L~1554, and consequently the amount of CO depletion at the
center of B68 is about two orders of magnitude lower than at the center
of L1544.  The intensity of the IRSF incident on B68 can be estimated
as follows.  We first estimate the external UV radiation field based on
the 90 \mic \ data of ~\citet{wtak02} and the conversion between FIR
intensity and incident field found by \citet{bp88}.  Another approach
is to use the 7\mic \ intensity from~\citet{bap00}  and the results of
\citet{bra96} to infer  $G_0$. Combining these, we conclude that the
radiation field incident on B68 is $\sim 2.5$ times larger that the
standard ISRF.

\begin{figure}
\resizebox{\hsize}{!}{\includegraphics{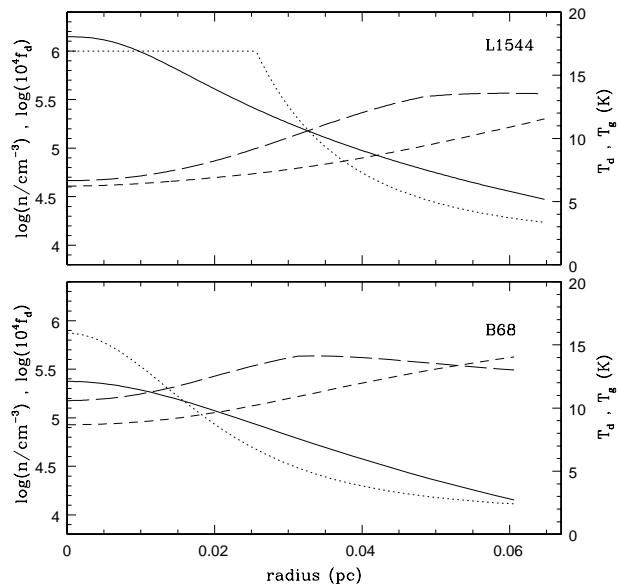}}
\caption{{\em Upper panel:} inferred gas ({\it long-dashed curve}\/) and dust
temperature dependence ({\it short-dashed curve}\/) 
for the model of L1544 discussed
in the text (with $G_0=1$). The assumed density distribution from
Eq.~(\ref{l1544den}) is shown for comparison ({\it solid curve}).
The assumed CO depletion factor $f_{\rm d}$ (multiplied
by $10^4$, {\it dotted curve}\/) is also shown.  {\em Lower panel:\/}
same as above, for the model of B68 (with $G_0=2.5$).  One sees that
for $n >3\times 10^4$ \percc\ the
gas temperature follows the dust temperature and decreases towards the
core center. Outside this rigion, gas and dust are decoupled and the
gas heating is determined solely by cosmic ray ionization.}
\label{fig_l1544}
\end{figure}

In Fig.~\ref{fig_l1544}, we show the inferred gas temperature
dependence for L1544 and B68, computed with cosmic ray ionization rate
$\zeta =1.3\times 10^{-17}$ s$^{-1}$ and accomodation coefficient
$=0.3$.  The assumed CO depletion factor (cut off at a value of 100) is
shown for comparison. In the inner part of both cores (densities above
$3\times 10^4$ \percc ), the gas temperature is coupled to that of the
dust and thus decreases gradually towards the core center. This
coupling is more effective in the case of L1544, which has a higher
central density.  The gas is slightly hotter than the dust however due
to cosmic ray heating and this difference increases as the density
decreases.  The CO depletion appears  to have only small effects on the
gas temperature profile because while at high density, gas--dust
coupling dominates, at densities below $10^5$ \percc, the depletion is
not very large (and the optically thick nature of CO cooling causes the
dependence of CO cooling on abundance to be minor).  We draw the
conclusion from this study that in the depleted core nucleus (densities
above $10^5$ \percc) where many molecular species may be unobservable
because they are in solid form, the gas temperature may be expected to
be close to that of the dust. In outer regions where the density is
below $10^5$ \percc \ and where shielding from the external radiation
field is much less, the gas and dust temperatures become uncoupled and
one may find gas temperatures lower than that of the dust.

It is worth pointing out that the scenario depicted in
Fig.~\ref{fig_l1544} yields gas temperatures in conflict with
observation.  \citet{tmc02} find based on their NH$_3$ observations a
temperature of 8.7~K throughout the L1544 core with no evidence for
density gradients (40\arcsec\ HPBW equivalent to 0.03 parsec), whereas
\citet{bal95} give $T_{\rm g}= 16$~K for B68, also derived from NH$_3$
data. These are in rough agreement with the values in Fig.~\ref{fig_l1544}
though the L1544 data do not show evidence for increasing temperature 
with radius as predicted. 

Another conclusion that one can draw from Fig.~\ref{fig_l1544} is that
the gas temperature and hence pressure in the high density core nucleus
depends somewhat on the external radiation field even for cores of
extremely high visual extinction. Thus one can expect cores in Taurus
for example (where the incident radiation field appears to be average)
to have different characteristics than Ophiuchus cores where the
radiation field appears in general to be an order of magnitude higher.
Of course, there are other effects such as the external pressure which
may differ in the two cases also but we stress that the radiation field
alone may cause core temperatures and hence accretion rates to be
higher in Ophiuchus than in Taurus.

\section{Self consistent hydrostatic equilibrium results}

The equation of hydrostatic equilibrium for a spherically symmetric
cloud is
\be
\frac{dp}{dr}+\rho\frac{dV}{dr}=0,
\ee
where $p$ is the gas pressure and $V$ the gravitational potential.
The latter satisfies Poisson's equation, which in spherical symmetry
reads
\be
\frac{1}{r^2}\frac{d}{dr}\left(r^2\frac{dV}{dr}\right)=4\pi G\rho.
\ee
The gas pressure $p$, neglecting the contribution of
turbulent pressure, is
\be
p=\frac{\rho}{\mu m_H}k T_{\rm g},
\ee
where $T_{\rm g}$ is the gas temperature, computed as described in
Sect.~3, and $\mu=2.33$ the mean molecular weight.

\begin{figure}
\resizebox{\hsize}{!}{\includegraphics{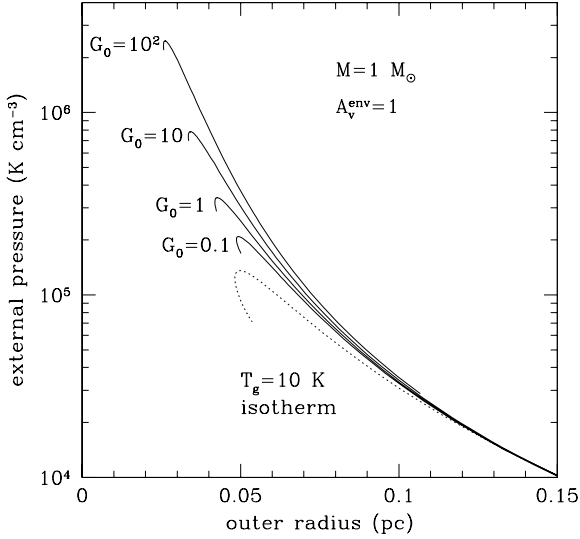}}
\caption{Sequences of model clouds with $M=1$~\msun\ and $A_{\rm V}^{\rm
env}=1$ in the radius--pressure diagram, for various scaling factors
$G_0$ of the ISRF ({\it solid lines}). For comparison, the {\it dotted
line} shows a Bonnor-Ebert isotherm with $T_{\rm g}=10$~K for the same
cloud mass.}
\label{fig_m1}
\end{figure}

The behavior of our model clouds with respect to an external
compression is qualitatively similar to that of isothermal spheres.
Fig.~\ref{fig_m1} shows the locus in the radius-external pressure
diagram of cloud models with $M=1$~\msun, for different values of the
ISRF intensity $G_0$.  For large radii the cloud follows closely the
$T_{\rm g}=10$~K Bonnor-Ebert isotherm, independent of the radiation
field (the effect of gas-dust coupling is negligible for these diffuse
configurations).  At higher values of the external pressure the cloud
radius is smaller and the central (and average) density becomes larger.
As a result, the central gas temperature increases and the sequence of
equilibria departs from the Bonnor-Ebert isotherm. Eventually, a
critical value of the pressure is reached above which no equilibria
exist, and the curve turns down.  Past this critical point, equilibria
are characterized by an increasing central density and density
contrast, and are unstable (see following Section). For clarity, in
Fig.~\ref{fig_m1} we truncate the curves of equilibrium slightly after
the critical point.

\begin{figure}
\resizebox{\hsize}{!}{\includegraphics{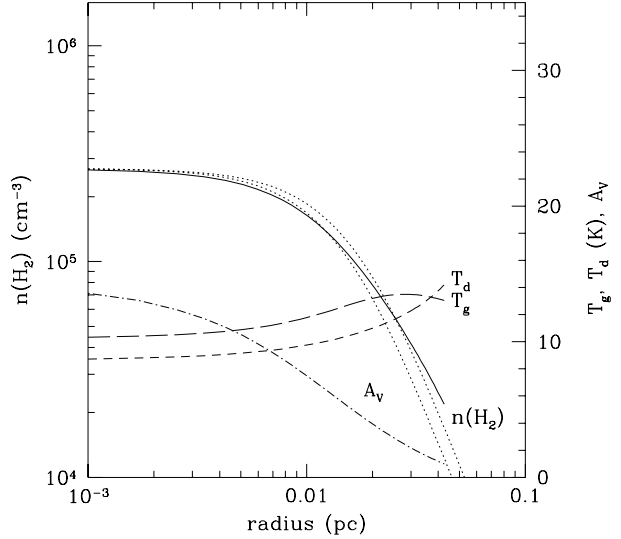}}
\caption{Density of H$_2$ ({\it solid line}), temperature of the dust
({\it short-dashed line}), temperature of the gas ({\it long-dashed
line}), and visual extinction ({\it dash-dotted line}) as function of
radius for the marginally stable clouds with $M=1$~\msun and
$G_0=1$. The {\it dotted lines} show for comparison the
density profile of an isothermal sphere with the same values of $n_{\rm c}$ 
and $T=T_{\rm g,c}$ or $T_{\rm g,b}$. The visual extinction
at the cloud's center is $A_{\rm V,c}=14$.}
\label{fig_m1g1crit}
\end{figure}

Fig.~\ref{fig_m1g1crit} shows the density of H$_2$, the temperature of
the gas and the temperature of the dust as function of radius for the
marginally stable clouds of mass $M=1$~\msun\ with $G_0=1$ (central
H$_2$ density $n_{\rm c} =2.7\times 10^5$~cm$^{-3}$).  The density
profile is very close to that of an isothermal sphere with the same
values of the central density, shown for comparison in
Fig.~\ref{fig_m1g1crit} for temperatures equal to the central and
boundary temperature of the model. The gas temperature decreases from
the outer regions to the center of the cloud. For marginally stable
clouds with $M=1$~\msun the central density is sufficient to ensure a
modest gas-dust coupling, but the gas temperature is mainly determined
by the balance of cosmic-ray heating and gas cooling.

\begin{figure}
\resizebox{\hsize}{!}{\includegraphics{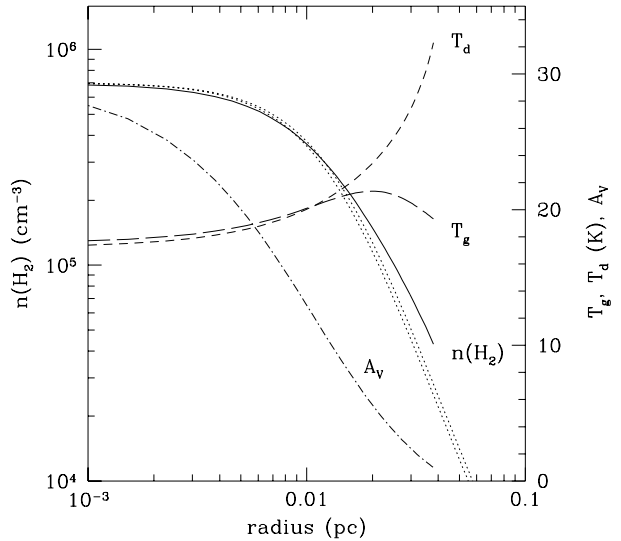}}
\caption{Same as in Fig.~\ref{fig_m1g1crit}, but for
a marginally stable cloud bounded by an external pressure
$p_{\rm ext}=1\times 10^6$~K~cm$^{-3}$ and with $G_0=100$.
The mass of the cloud is $M=1.6$~\msun\ and the extinction
from the center to the edge of the cloud is $A_{\rm V}=29$.}
\label{fig_g100pre1e6crit}
\end{figure}

Fig.~\ref{fig_g100pre1e6crit} shows the H$_2$ density, the gas and dust
temperatures and the extinction profile of the marginally stable cloud
in thermal equilibrium with an ISRF with $G_0=100$ and bounded by an
external pressure $p_{\rm ext}=1\times 10^6$~K~cm$^{-3}$. These roughly
correspond to the structures observed by \citet{jfmm2001} in the $\rho$Oph
and Orion clouds. For comparison, the figure also shows the density
profiles of marginally stable Bonnor-Ebert spheres with the same
central density and uniform temperature equal to the gas temperature at
the center and at the boundary of the cloud.

In Tables~\ref{tab_m1} and \ref{tab_m5} we list the values of central
H$_2$ density, gas and dust temperature, density contrast, and optical
extinction at the center for marginally stable clouds with $M=1$ and
5~\msun and various scaling factors for the ISRF. In these models 
we have varied $p_{\rm ext}$ until we find the marginally stable 
model.

A comparison of the results for the two different masses is
instructive. Small mass clouds ($M\simeq 1$~\msun) exposed to the
average ISRF behave like isothermal spheres with $T_{\rm g}\simeq
10$~K, and are characterized by a maximum density contrast and critical
central density very close to those of the corresponding Bonnor-Ebert
values. Increasing the ISRF ($G_0=10$) has the effect of raising the
dust temperature, and of increasing slightly the gas temperature in the
central regions. This increase in the central gas temperature causes a
higher central density in the marginally stable core and this, in turn,
increases the efficiency of dust-gas coupling.  A further increase of
the ISRF ($G_0=100$) establishes the approximate equality of the gas
and dust temperatures at the center.  Thus, the fact that small mass
clouds are dynamically stable for larger values of the central density
than their more massive counterparts (see Eq.~\ref{ncrit}), has the
consequence that they are ``stabilized'' by an external ISRF simply
because they are able to become warmer if the dust does.  Their gas
temperature profiles, however, remain quite uniform, and therefore the
maximum density contrast cannot deviate significantly from the
Bonnor-Ebert value $\rho_{\rm c}/\rho_{\rm b}\simeq 13.98$, as shown in
Table~\ref{tab_m1}.

A different behaviour characterizes more massive clouds ($M\simeq
5$~\msun), as they become dynamically unstable before an efficient
gas-dust coupling can be established, for any value of the intensity of
the ISRF (see Table~\ref{tab_m5}).  Their central extinction is low
($A_{\rm V,c}\simeq 3$--5), and hence they do not correspond to ``real
observed cores''. Nevertheless, it is interesting that they can develop
significant inward temperature gradients, because molecular cooling is
less efficient than cosmic-ray heating in the central regions,
characterized by densities $n\simeq 10^4$--$10^5$~cm$^{-3}$ (where
$\Lambda_{\rm g}\propto n^{0.5}$) than in the outer parts where
$n\simeq 10^3$ (and $\Lambda_{\rm g}\propto n$, as cosmic-ray heating).
An inward temperature gradient stabilizes the cloud (see Sect.~6), and
allows large deviations in the maximum density contrast $\rho_{\rm
c}/\rho_{\rm b}$, up to 30--40, as shown in Table~\ref{tab_m5}. Notice
that any molecular depletion effect, ignored in these calculation,
would make the central temperature even higher, and thus increase the
slope of the temperature gradient.

\begin{figure}
\resizebox{\hsize}{!}{\includegraphics{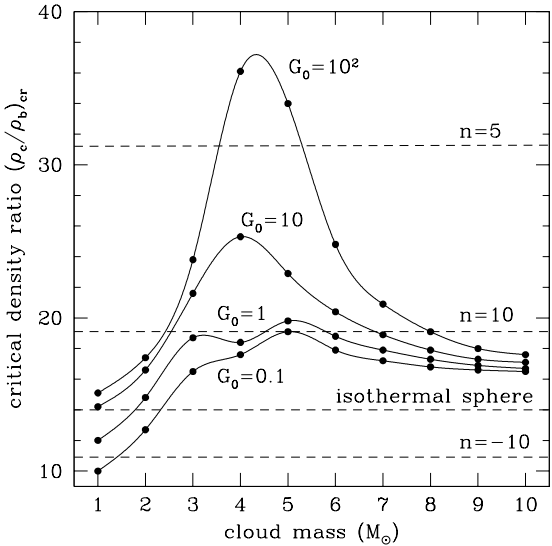}}
\caption{Value of the central-to-boundary density ratio $(\rho_{\rm
c}/\rho_{\rm b})_{\rm cr}$ as function of the cloud's mass for different
scaling factors $G_0$ of the ISRF ({\it dots and solid lines}).  The {\it
dashed lines} show the corresponding values of $(\rho_{\rm c}/\rho_{\rm
b})_{\rm cr}=13.98$ for marginally stable isothermal and polytropic spheres
of index $n$.}
\label{fig_rapp}
\end{figure}

In Fig.~\ref{fig_rapp} we summarize our results for the maximum density
contrast $\rho_{\rm c}/\rho_{\rm b}$ allowed for clouds in the mass
range $M=1$--10~\msun. Again $p_{\rm ext}$ has been varied until the
marginally stable model is found. Deviations of $\rho_{\rm c}/\rho_{\rm
b}$ from the Bonnor-Ebert value 13.98 indicate deviations from a
condition of uniform gas temperature in the cloud. As anticipated in
the above discussion, the largest deviations from isothermality are
obtained for masses $M\simeq 4$--6~\msun.  Marginally stable clouds of
small mass ($M\simless 2$~\msun) are characterized by relatively large
values of the central density, and by a sufficiently good coupling of
gas and dust: they are uniformly warmer or colder according to the
intensity of the external ISRF, but because of this coupling they do
not develop significant temperature gradients, and therefore behave
essentially as isothermal spheres.  Marginally stable high-mass clouds
($M\simgreat 8$~\msun) also are similar to isothermal spheres, but for
a different reason: they are characterized by relatively low values of
the central density ($n_{\rm c} \simeq 10^3$~cm$^{-3}$) and also have
low $A_{\rm V,c}$ (unlike observed cores), which implies an uniform
value of the gas temperature since both cosmic-ray heating and
radiative gas cooling depend approximately linearly on density (and
gas-dust coupling plays no role at these densities).  In the
intermediate mass range, both because of the different dependence of
heating and cooling rates from density and partial gas-dust coupling,
the deviations from uniform gas temperature are larger, and the maximum
density contrast $\rho_{\rm c}/\rho_{\rm b}$ has a peak for $M\simeq
4$--5~\msun. Fig.~\ref{fig_rapp} also shows, for comparison, the values
of the density contrast for marginally stable polytropic spheres of
index $n$, discussed in the next Section.

\begin{table}
\caption{\sc Properties of marginally stable clouds with $M=1$~\msun}
\vspace{1em}
\begin{tabular}{llllll}
\hline
$G_0$  & $n_{\rm c}$ & $\rho_{\rm c}/\rho_{\rm b}$ & $T_{\rm g,c}$ & $T_{\rm d,c}$ & $A_{\rm V,c}$ \\
       & (\percc)          &     & (K)           & (K)           &               \\
\hline
0.1    & $1.5\times 10^5$  & 10.0 & 9.2  &  6.2 &  9  \\
1      & $2.7\times 10^5$  & 12.0 & 8.7  & 10.3 & 14  \\
10     & $5.8\times 10^5$  & 14.4 & 12.7 & 12.0 & 22  \\
$10^2$ & $1.3\times 10^6$  & 15.1 & 17.0 & 16.8 & 38  \\
\hline
\end{tabular}
\label{tab_m1}
\end{table}

\begin{table}
\caption{\sc Properties of marginally stable clouds with $M=5$~\msun}
\vspace{1em}
\begin{tabular}{lllllll}
\hline
$G_0$  & $n_{\rm c}$ & $\rho_{\rm c}/\rho_{\rm b}$ & $T _{\rm g,c}$ & $T_{\rm d,c}$ & $A_{\rm V,c}$ \\
       & (\percc)  &    & (K)           & (K)           &          &         \\
\hline
0.1    & $9.9\times 10^3$  & 19.2 & 10.0 & 9.4  & 3  \\
1      & $1.1\times 10^4$  & 20.0 & 10.0 & 14.2 & 3  \\
10     & $1.5\times 10^4$  & 24.2 & 10.0 & 21.4 & 4  \\
$10^2$ & $2.4\times 10^4$  & 34.4 & 10.5 & 32.3 & 5  \\
\hline
\end{tabular}
\label{tab_m5}
\end{table}

\section{The effect of the external radiation field}

The presence of an external ISRF affects the stability of clouds bounded
by an external pressure. In the first place, it makes the dust (and
therefore the gas) warmer on average, thus increasing the critical
mass for gravitational instability (proportional to $T_{\rm g}^2$);
in the second place, it enforces a temperature gradient in the cloud
which modifies the value of the classic Bonnor-Ebert critical mass.

The latter point is best investigated by the so-called {\it static
method} originally devised by \citet{z63} for white dwarfs and neutron
stars \citep[see][]{t78}.  This method allows one to perform a
stability analysis for axisymmetric motions on the basis of the
properties of the equilibrium models only.  The static criterion
asserts that a turnover in the $M(\rho_{\rm c})$ curves at fixed
$P_{\rm ext}$, occurring at say $\rho_c=\rho_c^{\rm cr}(p_{\rm ext})$,
marks the onset of dynamical instability to radial motions.  The
condition $(\partial M/\partial \rho_{\rm c})_{p_{\rm ext}}=0$ is both
a sufficient and necessary condition for a stability transition:
equilibria with $\rho_{\rm c} < \rho_{\rm c}^{\rm cr}(p_{\rm ext})$ are
stable, while equilibria with $\rho_{\rm c} > \rho_{\rm c}^{\rm
cr}(p_{\rm ext})$ are unstable to at least one normal mode.

\begin{figure}
\resizebox{\hsize}{!}{\includegraphics{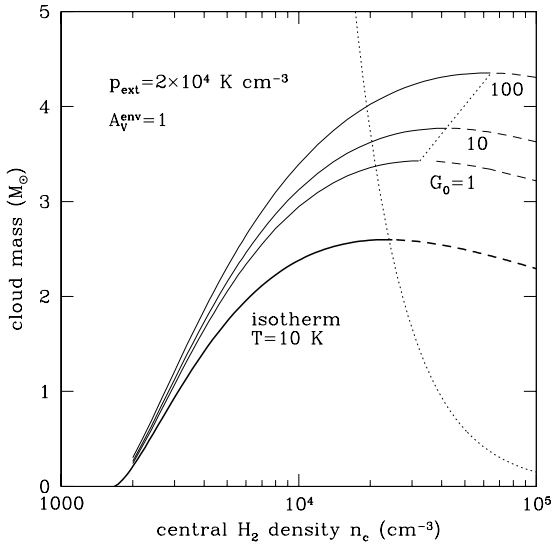}}
\caption{Instability in model cloud cores for isothermal and
non-isothermal models. The cloud's mass is plotted as function of the
central H$_2$ density $n_{\rm c}$ for $p_{\rm ext}=2\times
10^4$~K~cm$^{-3}$ ({\it solid curves}, stable equilibria; {\it dashed
curves}, unstable equilibria).  Instability sets in at the maximum 
of $M(n_{\rm c})$.  The curves are labelled with the values
of $G_0$. The {\it dotted lines} separate the regions of stability and
instability for this value of $p_{\rm ext}$ and arbitrary values of the
temperature (in the isothermal case) and $G_0$ (in the non-isothermal
case). The {\it thick curve} show the mass versus central density
relation for an isothermal sphere with $T_{\rm g}=10$~K.}
\label{fig_static}
\end{figure}

In Fig.~\ref{fig_static} we show how the intensity of the external ISRF
may affect the stability of a cloud bounded by a given external
pressure. For this particular example, we have chosen $p_{\rm
ext}=2\times 10^4$~K~cm$^{-3}$ \citep[][]{mk99}.  The three curves
labelled with $G_0=1$, 10 and 100 show the mass versus central H$_2$
density for externally heated clouds, compared with the similar
quantities for isothermal spheres.  According to the static method,
continuous (dashed) lines represent stable (unstable) equilibria.  The
dotted lines indicates the boundaries of the stability domain for
isothermal spheres with arbitrary temperatures (only the case $T_{\rm
gas}=10$~K is shown) and non-isothermal spheres in thermal equilibrium
with the external ISRF.  The figure shows that for a given external
pressure the condition of thermal equilibrium allows the existence of
stable cloud configurations for values of mass and central density not
permitted under the condition of uniform gas temperature.

Although not very large, this effect may resolve in part the problem
posed by objects like B68 which are well fitted by slightly unstable
Bonnor-Ebert spheres (magnetic fields are of course another
possibility). For example, for the three marginally stable
non-isothermal clouds of Fig.~\ref{fig_static} the central-to-boundary
density ratio are $(\rho_{\rm c}/\rho_{\rm b})=19$, 25 and 39, for
$G_0=1$, 10 and 100, compared with the corresponding ratio 13.98 for
isothermal spheres, and 16.5 inferred for B68~\citep{al01}.

\subsection{Comparison with polytropic models}

From the point of view of the stability properties and the maximum
density contrast, the models shown in this paper have a behavior
similar to polytropes with $n \gg 1$ (see Fig.~\ref{fig_m1g1crit}).
Increasing the intensity of the ISRF produces an increase in the dust
temperature which is felt by the gas only in the central parts of the
cloud, where the coupling of dust and gas is more efficient. As a result,
the central gas temperature $T_{\rm g,c}$ increases slightly, whereas
the gas temperature at the boundary $T_{\rm g,b}$ remains unchanged
($T_{\rm g,b}\simeq 10$~K), since for densities $n\simeq 10^3$~cm$^{-3}$
both the gas cooling rate and the the cosmic-ray heating rate are roughly
proportional to $n$.  Thus, one can extend Eq.~(\ref{mbe}) to polytropic
clouds bounded by the same external pressure and characterized by the
same thermal sound speed $a_{\rm b}$ at the boundary, and write
\be 
M_{\rm cr}\simeq \alpha_n
\frac{a_{\rm b}^4}{\sqrt{G^3p_{\rm ext}}}.  
\ee 

Table~\ref{tab_poly} shows the value of $\alpha_n$ and the density
contrast $\rho_{\rm c}/\rho_{\rm b}$ for clouds with different
polytropic index $n$. For decreasing positive $n$, the coefficient
$\alpha_n$ increases, indicating a larger critical mass than an
isothermal sphere bounded by the same external pressure and with the
same temperature at the boundary.  For $n=6$, for example, the critical
mass is expected to be a factor $2.06/1.18\simeq 1.75$ larger than
critical Bonnor-Ebert mass for the same gas temperature at the
boundary ($M_{\rm BE}\simeq 2.6$~\msun\ for $T_{\rm g}=10$~K, see
Eq.~\ref{mbe}). 

\begin{table}
\caption{\sc Critical mass and density contrast 
as function of polytropic index}
\vspace{1em}
\begin{tabular}{lll}
\hline
$n$     & $\alpha_n$ & $\rho_{\rm c}/\rho_{\rm b}$ \\
\hline
3.5       & 3.37      & 88   \\
4         & 2.85      & 50   \\
5         & 2.33      & 32   \\
10        & 1.64      & 19   \\
20        & 1.39      & 15   \\
$\infty$  & 1.18      & 14   \\
\hline
\end{tabular}
\label{tab_poly}
\end{table}

\section{Discussion and Conclusions}

We have examined the gas temperature distribution to be expected in
interstellar pre--protostellar cores heated by the external ISRF. We
find that when (as in observed cores), the central density exceeds
$3 \times 10^4$ \percc , there is coupling between the gas and dust
temperatures and hence the gas temperature (like the dust) decreases with
decreasing radius.  At larger radii and smaller densities, the dust and
gas decouple and the gas temperature may (for low external pressures as
in the Taurus cloud) decrease towards the values expected for heating
by galactic cosmic rays of around 10~K. The region where gas and dust
temperatures are coupled is somewhat interior to the region where CO
is highly depleted.  However, CO depletion does not seem greatly to
affect the temperature distribution mainly because cooling by gas--grain
collisions becomes dominant.

The observed values of the temperature of around 10~K in many
pre-stellar cores allow limits to be placed on the cosmic ray
ionization rate similar to the standard value of order $10^{-17}$
s$^{-1}$ based on the measured cosmic ray flux. The fact that measured
gas temperatures in cores show so little spatial variation
\citep{bm89,tmc02} suggests to us that the observed thermometers mainly
trace layers  of moderate depletion where gas--grain coupling is not
playing an important role.

We have also examined the consequences of such a temperature
distribution for the density distribution in hydrostatic equilibrium
cores. The changes caused by the ``real temperature distribution'' are
minor and the characteristics of a ``marginally stable Bonnor-Ebert
sphere'' are similar to those in the isothermal case. 

An interesting point  which emerges from these calculations is that the
temperature in the core nucleus is sensitive to the external radiation
field (because the core nucleus as a rule is heated by small particle
MIR emission from the borders of the surrounding cloud). This is in
particular the case for cores with high external pressure (and hence
high density) such as those studied by ~\citet{jfmm2001,jwm2000} in the
Orion B and $\rho$Oph clouds where $p_{\rm ext}$ appears to be above
$10^6$ \percc K. In these cases, we expect the dust temperature to
follow the gas temperature and thus the behavior should be roughly like
a negative index polytrope. The external radiation fields are also
higher in Orion and Ophiuchus than in Taurus and thus the core masses
may also rise somewhat. For constant external pressure, the mass of the
marginally stable Bonnor-Ebert sphere rises with about the 0.35 power
of the external radiation field. It would be useful to have direct
temperature estimates to confirm these expectations.

Another result of this study is that high mass thermally supported
cores are incompatible with high column density. Thus it is difficult
to imagine observed high column density (greater than $10^{22}$ \cmsq)
high mass (greater than 10~\msun) cores going through a series of
quasi-equilibrium states en route to collapse. Hence we suspect that
higher mass cores  are either not stable or have other (magnetic) means
of support.  These other means of support become apparent
observationally both because observed cores show large departures from
spherical symmetry and also because the observed line widths in most
cores  can only be explained in terms of turbulence and often of
supersonic turbulence.

Thus thermally supported cores may be the exception rather than the
rule. It is significant nonetheless that cores in regions such as
Taurus where predominantly low mass star formation is taking place have
in general line profiles showing a large component of thermal
broadening. This suggests that where (as in Taurus), the pivotal state
may be a core with predominantly thermal support, only low mass stars
are likely to form.  In clouds such as Ophiuchus and Orion with higher
radiation fields and pressures and with cores having predominantly
non--thermal support, higher mass stars may become possible and star
formation may take another course.

It is also significant that occasionally, one finds cases like B68
where the data are consistent with hydrostatic equilibrium (marginally
stable) and pure thermal support. In fact, we find that when one takes
the gas temperature dependence into account, B68 is marginally
stable. But irrespective of whether this is true or not, B68 gives
every sign of being close to the pivotal state from which protostellar
collapse will commence.

\acknowledgements

It is a pleasure to thank A. Natta for constructive criticism and
Jens Kauffman for useful comments on the manuscript. We acknowledge 
financial support from the EC Research Training Network ``The Formation 
and Evolution of Young Stellar Clusters'' and grant COFIN-2000.
JG acknowledges support from scolarship SFRH/BD/6108/2001 awarded by 
the Portuguese Ministry of Education.

\bibliography{}

\bibliographystyle{aa}

\end{document}